\documentclass[aps,prl,twocolumn,longbibliography,10pt]{revtex4-2}
\usepackage{siunitx}
\usepackage{isotope}
\usepackage{times}
\usepackage{braket}
\usepackage{graphicx,lipsum}
\usepackage{here}

\usepackage[T1]{fontenc}

\begin{document} 


\title{A quantum-network register assembled with optical tweezers in an optical cavity}

\author{Lukas Hartung$^{1}$}
\email[To whom correspondence should be addressed. Email: ]{lukas.hartung@mpq.mpg.de}
\author{Matthias Seubert$^{1}$}
\author{Stephan Welte$^{1,2}$}
\author{Emanuele Distante$^{1}$}
\altaffiliation[\textit{Present address:}]{ ICFO-Institut de Ciencies Fotoniques, The Barcelona Institute of Science and Technology, 08860 Castelldefels (Barcelona), Spain.}
\author{ Gerhard Rempe$^{1}$}
\affiliation{$^{1}$Max-Planck-Institut f\"ur Quantenoptik, Hans-Kopfermann-Straße 1, 85748 Garching, Germany}
\affiliation{$^{2}$Institute for Quantum Electronics, ETH Z\"urich, Otto-Stern-Weg 1, 8093 Zürich, Switzerland}


\begin{abstract}

\noindent Quantum computation and quantum communication are expected to provide users with capabilities inaccessible by classical physics. However, scalability to larger systems with many qubits is challenging. One solution is to develop a quantum network consisting of small-scale quantum registers containing computation qubits that are reversibly interfaced to communication qubits. Here we report on a register that uses both optical tweezers and optical lattices to deterministically assemble a two-dimensional array of atoms in an optical cavity. Harnessing a single-atom addressing beam, we stimulate the emission of a photon from each atom and demonstrate multiplexed atom-photon entanglement with a generation-to-detection efficiency approaching 90\%. Combined with cavity-mediated quantum logic, our approach provides a possible route to distributed quantum information processing.
\end{abstract}
  
  \maketitle

\noindent Quantum networks \cite{Kimble2008,wehner_quantum_2018} offer potential for numerous applications, including secure communication \cite{Pirandola2020}, distributed quantum computing \cite{goos_distributed_2003,monroe_large-scale_2014,daiss_quantum-logic_2021}, precision sensing \cite{Gottesman2012}, and clock synchronization \cite{Komar2014}. The envisioned architecture comprises network nodes with stationary qubits for processing quantum information, connected via optical fibers for the exchange of photonic qubits. However, optical losses and unavoidable errors introduce substantial challenges in practical implementations. The solution is using a scalable multi-qubit register per node: this allows overcoming losses with multiplexed communication protocols \cite{Sangouard2011,krutyanskiy_multimode_2023} including repeaters \cite{Sangouard2011,luong_overcoming_2016,langenfeld_quantum_2021,krutyanskiy2023}, as well as increasing fidelities through entanglement distillation \cite{Kalb2017} and quantum error correction \cite{Roffe2019}.

The goal then is to create a register of individually controllable qubits and couple each of them to a photonic channel for network connectivity \cite{Covey2023}. We address this twofold challenge by combining cavity quantum electrodynamics (QED) \cite{Reiserer2015} with the atom-based technologies of optical lattices and optical tweezers \cite{schlosser_sub-poissonian_2001}. Specifically, the cavity serves as an atom-photon quantum interface, the lattice provides minimally perturbing atomic localization, and the tweezers are used to assemble on demand any desired atomic configuration by means of tightly focused laser beams that can be maneuvered at will \cite{kim_situ_2016,barredo_atom-by-atom_2016,endres_atom-by-atom_2016}. While large arrays of tweezers can realize registers with hundreds of qubits \cite{Ebadi2021,scholl_quantum_2021}, no experiment has been performed in the desired network setting, arguably because the cavity structure restricts the optical access required for the tweezers. In fact, despite long-standing efforts, only a few examples of cavity-coupled atoms trapped in tweezers have been reported \cite{Dordevic2021,deist_mid-circuit_2022,Liu2023}. All of them lack individual control over the atomic qubit, a necessary requirement for applications in a quantum network \cite{DiVincenzo2000}.

With an array of optical tweezers and the intra-cavity optical lattice, we achieve individual control over a set of atomic qubits in a network setting. We create one- and two-dimensional registers of up to $n{=}6$ atoms and address each atom separately to generate atom-photon entanglement (APE). We observe that the APE fidelity remains constant for up to six qubits in the register, indicating that scalability to larger system sizes is possible. By implementing a multiplexing scheme, we generate APE with a probability exceeding $\sim$\,$\qty{97}{\percent}$ per attempt, an important step towards the deterministic distribution of entanglement across networks. All this is achieved by exploiting the advantages of a Fabry-P\'{e}rot cavity that allows for optical access in a plane perpendicular to its symmetry axis.

\textbf{Experimental setup.} Figure 1A schematically shows our setup to prepare and individually control two-dimensional arrays of in our case \isotope[87]{Rb} atoms at the center of the cavity. Laser beams propagating along the $z$-axis are focused at the cavity plane to $\qty{1.4}{\micro \metre}$ waist radius by an objective (NA=0.4), and their positions are precisely controlled using two (only one shown) two-dimensional acousto-optic deflectors (2D-AODs). This enables selective addressing of individual atoms across a broad wavelength range from $\qtyrange{770}{810} {\nano \metre}$, covering the $D_1$ and $D_2$ line of \isotope[87]{Rb}. Atom arrays are prepared using $\qty{797}{\nano \metre}$ light and driving one pair of AODs with multiple radio-frequency (RF) tones \cite{endres_atom-by-atom_2016}. This creates a static two-dimensional array of tweezers which are stochastically loaded with cold atoms from a magneto-optical trap. Atoms are cooled using pairs of counter-propagating laser beams aligned at 45 degrees in the $x,z$ plane (not shown) \cite{nusmann_vacuum-stimulated_2005}. By detecting the scattered cooling light using an electron-multiplying charge-coupled device (EMCCD) camera, we image the initial atoms positions. Based on the images captured, we use the second pair of AODs to create additional tweezers that move and arrange the atoms into ordered arrays within the static tweezers. We observe an increase of about four orders of magnitude in the success rate of preparing arrays of six atoms compared to stochastic loading (see supplement \cite{supp}), thus laying the foundation for loading many more atoms than just one or two as in previous experiments.

\begin{figure*}[htbp]
\centering
\includegraphics[width=2\columnwidth]{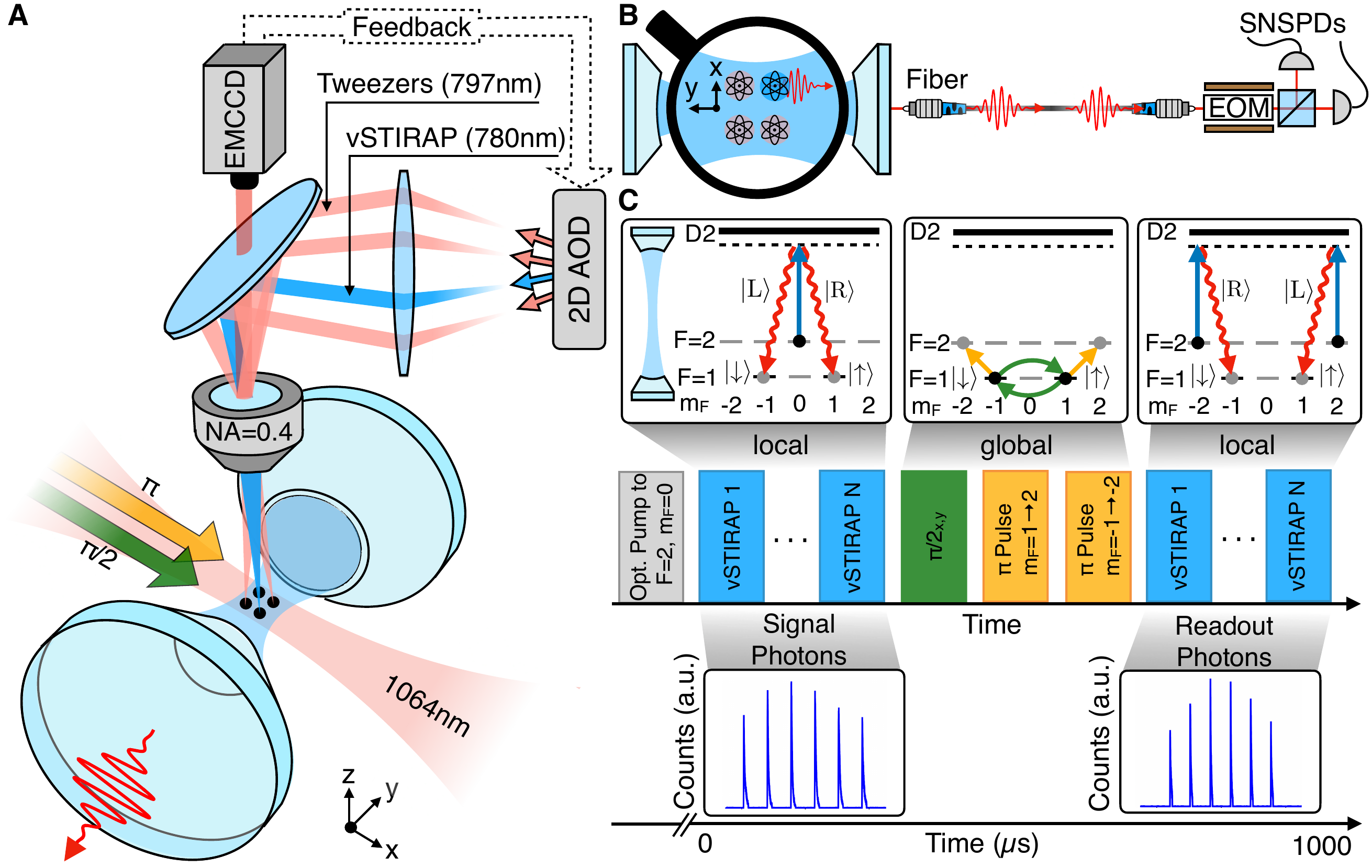}
\caption{\label{fig:single_tweezer}\textbf{Setup and experimental sequence. (A)} Tweezers and vSTIRAP lasers are steered via 2D-AODs onto individual atoms at the cavity center to generate and control two-dimensional atomic arrays. Atoms are imaged via an electron-multiplying charge-coupled device (EMCCD) camera, and feedback based on the camera pictures is applied to the 2D-AODs to prepare ordered atomic arrays. After the preparation, the atoms are transferred to a 2D lattice formed by a \qty{1064}{\nano \metre} standing wave trap and a \qty{770}{\nano \metre} intra-cavity standing wave trap. Global Raman lasers drive transitions between the ground states. \textbf{(B)} Photons emitted into the cavity mode predominantly exit via the outcoupling mirror and pass through an optical fiber and an electro-optical modulator (EOM), allowing to switch the detection basis of the photons. The photons are eventually detected with efficient superconducting nanowire detectors. \textbf{(C)} Protocol for APE generation. After optical pumping to $\ket{F{=}2, m_F{=}0}$, we apply sequentially $N$ local vSTIRAP pulses for photon generation (signal photons, red wiggly arrows). Global $\pi/2$ and $\pi$ pulses set the detection basis for the atom readout. Eventually, $N$ vSTIRAP pulses generate $N$ readout photons. The measured time trace in blue shows the signal and readout photons.}
\end{figure*}

Once the preparation is completed, the atoms are transferred from the tweezers to a two-dimensional optical lattice consisting of a red-detuned standing wave trap at \qty{1064}{\nano \metre} orthogonal to the cavity axis ($x$-axis in Fig. 1A) and an intra-cavity blue-detuned standing wave trap at \qty{770}{\nano \metre} ($y$-axis in Fig. 1A). During this process, the atom position may change by up to \qty{328}{\nano \metre}. This is negligible with regard to the typical atom separation of a few \qty{}{\micro \metre} given by the distance between the tweezers. Compared to trapping atoms in tweezers, the lattice ensures stronger atom confinement and overcomes limitations in array size imposed by finite laser power. Crucially, the lattice avoids the emergence of near-focus extraordinary polarization components, as observed in tightly focused optical tweezers \cite{Rosenfeldt2008,Thompson2013}. These components create large virtual magnetic fields, reducing the coherence of magnetically sensitive atomic qubits and posing challenges for entangling atoms and photons in a cavity. The \qty{770}{\nano \metre} light is also used to stabilize the length of the cavity, which is a high-finesse Fabry-P\'{e}rot resonator with unbalanced mirror transmissivities $T{=}\qty{4}{ppm}$ and $T{=}\qty{92}{ppm}$ (finesse $\mathcal{F}=\qty{61(2)e3}{}$). This ensures that light within the cavity exits predominantly through the higher-transmissivity outcoupling mirror. Photons exiting from the cavity pass through an optical fiber and are detected with a polarization-resolving detection setup (Fig. 1B) consisting of an electro-optic modulator (EOM), a set of waveplates (not shown), a polarizing beamsplitter (PBS), and two fiber-coupled superconducting nanowire single-photon detectors (SNSPDs). State initialization and coherent Raman transitions between the ground states are implemented with large laser beams aligned orthogonal to the cavity that illuminate the atoms simultaneously with homogeneous intensity. Throughout the entire experimental sequence, we apply a bias magnetic field aligned with the cavity axis, inducing a Zeeman splitting with a Larmor frequency of $\omega_L= 2\pi \times \qty{100}{\kilo \Hz}$. 

\begin{figure*}
\centering
\includegraphics[scale=1]{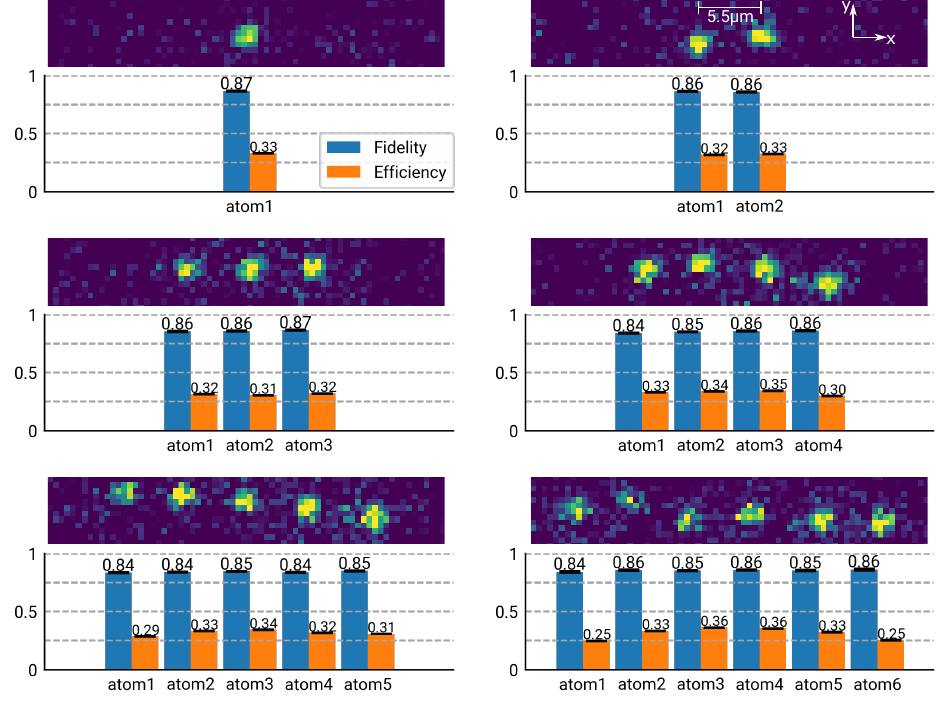}
\caption{\label{fig:FidEffVsNumOfAtoms}\textbf{Images of the atom arrays for up to $\mathbf{n=6}$ atoms with efficiency and fidelity for each atom.} The atoms are arranged along the $x$-axis, perpendicular to the cavity mode. The atoms are spaced $\qty{5.5}{\micro \metre}$ apart. From time to time, the atoms jump along the resonator axis within the blue standing wave trap ($y$-axis). By constantly evaluating the images and recalculating the necessary RF frequencies, we can follow the movement of the atoms with our addressing system. The APE fidelity $\mathcal{F}$ for each individual atom is independent of the number and the position of the atoms in the cavity. However, the efficiency of the photon emission process decreases for atoms displaced from the center of the cavity as the coupling $g$ decreases.
}
\end{figure*}

\quad \newline
\textbf{Quantum protocol}. Once the atomic array of $n$ atoms has been successfully prepared, we optically pump all atoms into the state $\ket{F{=}2, m_F{=}0}$ (Fig. 1C). We then illuminate the atoms one after the other via the addressing system with a $\qty{780}{\nano \metre}$ $\pi$-polarized light, red-detuned relative to the $D_2$ line. The resonator frequency is tuned such that, together with the $\pi$-light, the cavity stimulates off-resonant Raman scattering of a pump photon into the cavity mode, a process known as vSTIRAP \cite{Morin2019}. This transfers the respective atom into one of the two possible states $\ket{ \uparrow} = \ket{F{=}1, m_F{=}+1}$ or $\ket{ \downarrow} = \ket{F{=}1, m_F{=}-1}$, depending on the polarization of the photon, referred to as the signal photon. The signal photon and the atom are then in the maximally entangled Bell state
\begin{equation}
\label{eq:PsiPlus}
 \ket{\Psi^+}=\frac{1}{\sqrt{2}}\left(\ket{\uparrow, R}+\ket{\downarrow,L}\right),
\end{equation}
with $\ket{R}$/$\ket{L}$ denoting a right/left circularly polarized photon. By addressing the $n$ atoms sequentially, we ideally generate $n$ atom-photon entangled pairs in $n$ predefined temporal modes, as shown in the measured time trace of Fig. 1C.

We determine the entanglement fidelity $\mathcal{F}=\bra{\Psi^+}\rho\ket{\Psi^+}$ of the generated state with density matrix $\rho$ by measuring the correlations between the signal photon polarization and the internal atomic state in three different bases. The signal photons are measured using the polarization resolving setup described above (see Fig. 1B). To readout the atomic state, a single-qubit rotation using a two-photon Raman transition between $\ket{\uparrow}$ and $\ket{\downarrow}$ (green pulses in Fig. 1C) is applied to set the measurement basis (see supplement \cite{supp}). We then transfer $\ket{\uparrow}$ to $\ket{F{=}2,m_F{=}2}$ and $\ket{\downarrow}$ to $\ket{F{=}2,m_F{=}-2}$ with two additional Raman pulses (yellow in Fig. 1C). Finally, we apply atom-selective vSTIRAP pulses to map each atomic state onto a corresponding state of a readout photon. By measuring the readout photons in the circular basis, the internal atomic state can be uniquely determined as $\ket{R}$ implies $\ket{F{=}2,m_F{=}-2} $ and $\ket{L}$ implies $\ket{F{=}2,m_F{=}2}$. We use the EOM to rapidly switch the detection basis of the polarization resolving setup since the signal photons are measured along any desired polarization basis while the readout photons are observed along the circular basis. \newline
\quad \newline
\textbf{A multi-qubit register.} We measure the APE fidelity for an increasing number of atomic qubits. Fig. 2 illustrates the shapes and sizes of the prepared atom arrays and indicates the APE fidelity $\mathcal{F}$ and efficiency $\eta$ for each atom individually. Because of their finite temperature, the atoms do not always stay perfectly aligned along the cavity axis. Instead, they can hop between different lattice sites, causing variations in their position. However, we continuously monitor the position of the atoms by taking fluorescence pictures and dynamically adjust the radio-frequency signal supplying the AODs to address each atom precisely. 

When only one atom is prepared in the cavity mode, we measure a fidelity of $\mathcal{F}=\qty{86.6(5)}{\percent}$ and an efficiency of $\eta = \qty{33.2(2)}{\percent}$. The fidelity is mostly limited by state preparation and measurement errors (\qty{2.7(2)}{\percent}), and the Zeeman splitting of $\ket{\uparrow}$ and $\ket{\downarrow}$ states, which results in a frequency difference of the $\ket{R}$ and $\ket{L}$ components, inducing a phase chirp within the temporal mode (\qty{3.8}{\percent}). The latter can be minimized by reducing the photon temporal window considered (see supplement \cite{supp}). In addition, state preparation and readout (\qty{2.7(3)}{\percent}), decoherence (on average \qtyrange{1.5}{3.5}{\percent}, depending on the number of qubits), infidelities in the single-qubit rotations (\qty{2.5}{\percent}) and polarization settings ($<\qty{1}{\percent}$) reduce the fidelity further. The APE efficiency is given by $\eta=\xi P$ where $P$ is the intrinsic probability to generate a photon at the output of the cavity and $\mathcal{\xi}$ is the product of the state initialization efficiency in $\ket{F{=}2,m_F{=}0}$ (\qty{80(5)}{\percent}) and the overall transmission and detection efficiency (\qty{64(6)}{\percent}). The intrinsic photon generation probability depends on the cavity QED parameters as $P = (\kappa_{out}/\kappa) \times 2C/(2C+1)$, where $C=g^2/(2\kappa \gamma)$ is the cooperativity with $g$ the atom-cavity coupling constant for the relevant transition, $\kappa$ and $\kappa_{out}$ the total cavity field decay rate and the field decay rate through the outcoupling mirror, respectively, and $\gamma$ the atomic polarization decay rate. For our system $\left(g, \kappa, \kappa_{out},\gamma \right) = 2 \pi \times \left(\qty{5.0}{},\qty{2.5}{}, \qty{2.3}{},\qty{3}{}\right)\qty{}{\mega \Hz}$ resulting in $P=70\%$. Using $\xi = \qty{51(6)}{\percent}$, the expected efficiency is $\eta=\qty{36(4)}{\percent}$, comparable to the measured value.

When larger atom arrays are prepared, we measure the efficiency and the fidelity of the APE for each atom individually. For each array size, we collect a minimum of 3000 clicks per atom per detection basis and extract the average fidelity $\overline{\mathcal{F}}$. We observe that the fidelity does not depend on the size of the array (Fig. 3A), remaining constant at $\overline{\mathcal{F}}=\qty{85.5(4)}{\percent}$ for up to $n=6$ atoms. This proves minimal errors ($<0.7\%$) associated with increasing the size of the qubit register and indicates that larger registers can be realized without increasing the error rate. Conversely, the efficiency is not constant over the entire array of atoms aligned transversely to the cavity axis. It peaks near the cavity center and diminishes for outer positions, where the cavity field and, consequently, $C$ are lower due to the finite cavity mode size ($\qty{30}{\micro \metre}$ waist radius). We confirm this by measuring $\eta$ for a single atom placed at different positions along the $x$-axis (Fig. 3B). Our data are well represented by a model that includes the finite cavity mode size. The number of atoms along the $x$-axis is thus limited, with larger arrays possible for shorter inter-atomic distances $d_x$ along the $x$-axis or by shuffling atoms in and out of the cavity mode \cite{Nussmann2005,Bluvstein2022}. To determine the optimal $d_x$, we measured the average APE fidelity for two atoms within the cavity as a function of $d_x$, ranging from \qty{3.5}{\micro\metre} to \qty{16.5}{\micro\metre} (Fig. 3C). While no cross-talk is observed, with the fidelity remaining constant within a measurement error of $0.4\%$ (standard deviation), we note that for small inter-atomic distances, the array preparation is less efficient due to interference from neighboring tweezers \cite{Kaufmann2014}. We choose $d_x = \qty{5.5}{\micro\metre}$ between adjacent atoms as a compromise between efficient photon emission and atom preparation.

\begin{figure}
\includegraphics[width=\columnwidth]{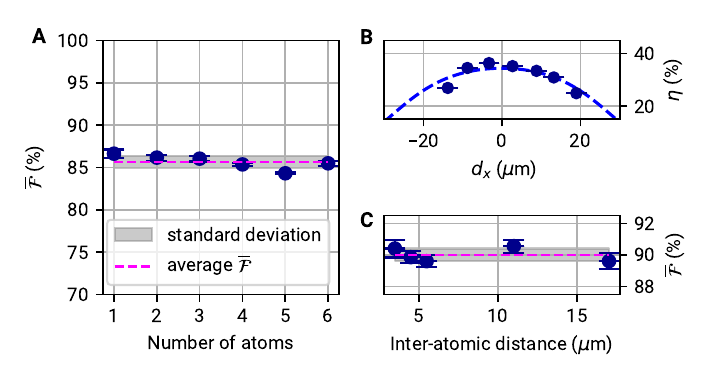}
\caption{\textbf{Average Fidelity $\overline{\mathcal{F}}$, cross-talk and position-dependent efficiency of APE. (A)} Average APE fidelity $\overline{\mathcal{F}}$ for $n$ atoms in the cavity. As more atoms are added, the fidelity remains constant at $\overline{\mathcal{F}}=\qty{85.5(4)}{\percent}$ within the standard deviation of \qty{0.7}{\percent}. \textbf{(B)} Efficiency $\eta$ of the photon emission process depending on the position of one atom along the $x$-axis (transversal to the cavity). As the atoms are positioned farther away from the center of the cavity, the coupling $g$ to the cavity and hence $\eta$ decreases. The dashed blue line is a fit of the data for the emission efficiency $\eta=\xi P$ with a model that includes the finite cavity waist $w_0$, the cavity-QED parameters $g, \kappa, \kappa_{out}, \gamma$ and uses only the overall transmission and detection efficiency  $\xi$ as free fit parameter. This yields $\xi=\qty{48.6(13)}{\percent}$, comparable with the measured value of $\qty{51(6)}{\percent}$. \textbf{(C)} Average APE fidelity for two atoms separated by $d_x$ along the $x$-axis. The fidelity remains constant at an average value of $\qty{90.0\pm 0.4}{\percent}$ (dashed line) indicating no cross-talk induced errors within the measurement error of $\qty{0.4}{\percent}$ (standard deviation).}
\end{figure}
\quad \newline
\noindent While the size of the arrays along the $x$-axis is limited, it is possible to extend it in two dimensions (see Fig. 4), positioning atoms along the cavity $y$-axis where the size is fundamentally constrained only by the cavity length of in our case $\qty{485}{\micro \meter}$. Our data consistently show that the fidelity remains constant as atoms are displaced along the cavity axis, albeit with a slight drop in efficiency. This results from our trapping scheme, which uses the intra-cavity blue-detuned trap to confine atoms at the nodes of the $\qty{770}{\nano \metre}$ standing wave. At the cavity center, the node of the $\qty{770}{\nano \metre}$ and the antinode of the $\qty{780}{\nano \metre}$ coincide, ensuring maximum atom-cavity coupling and thus $C$. However, the two standing waves acquire a different phase as we move along the cavity axis. Consequently, the effective $C$ for atoms trapped along the $y$-axis oscillates with a $\qty{32}{\micro \meter}$ beating pattern between the two wavelengths. This effect can be mitigated by selecting a trapping wavelength closer to $\qty{780}{\nano \metre}$. Although there is space for positioning many atoms along the cavity axis, currently, our setup is limited to two-row configurations due to the small (\qty{12}{\micro \metre}) beam waist of the \qty{1064}{\nano \metre} trap which limits the overall size of the optical lattice. A larger waist of the \qty{1064}{\nano \metre} trap, or multiple parallel standing-wave traps could be used to overcome this limitation. 
\begin{figure}
\includegraphics[width = \columnwidth]{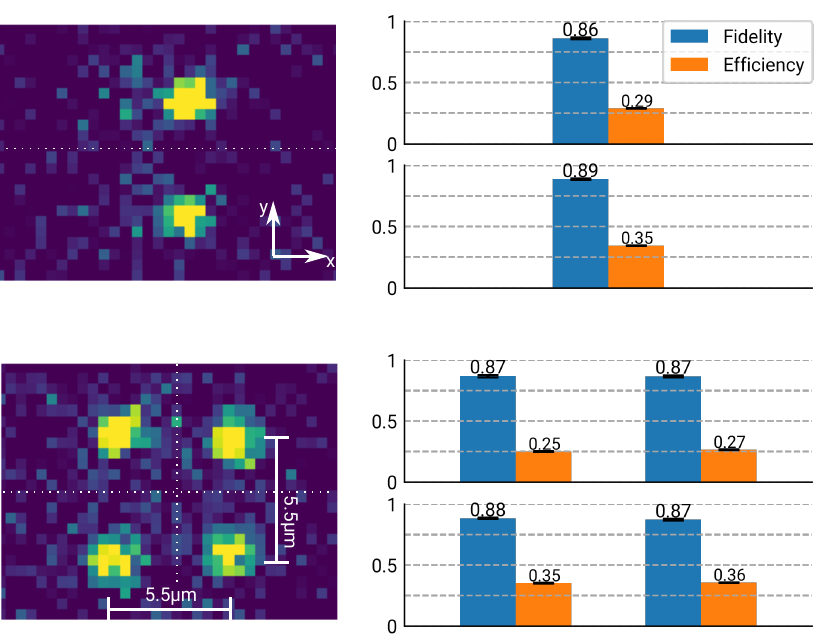}
\caption{\label{fig:2DArray} \textbf{Two-dimensional arrangement of atoms.} 
Upper part: Two atoms are arranged along the direction of the cavity mode, the $y$ direction in Fig. 1A. Lower part: Four atoms arranged as a $2\times2$ square along the $x$ and $y$ directions. On the right, the achieved fidelities and efficiencies of the atom-photon entangled states are presented. Our ability to arrange the atoms in two dimensions at constant fidelities and at the expected efficiencies indicates the scalability of the platform to larger system sizes, the maximum number of atoms being limited only by the mode volume of the cavity.
}
\end{figure}
\newline
\newline
\textbf{Multiplexed atom-photon entanglement generation with a quantum network register.} Despite a notable $\qty{33.2(2)}{\percent}$ probability of generating, propagating and detecting entanglement with a single atom, practical network applications demand pushing this probability as close to unity as possible. Since only one entangled state is needed to establish quantum communication between two nodes, using multiplexing schemes for registers of $n$ emitters provide a solution \cite{huie_multiplexed_2021,krutyanskiy_multimode_2023}. Given $\eta_i$ is the probability of detecting one photon from the $i$-th atom, the probability of detecting at least one entangled pair out of $n$ atoms is given by $\eta_{overall}= 1- \prod_{i=1}^n (1-\eta_i)$, approaching unity for large $n$. As each emitter can be addressed at different times, the communicating parties can distinguish which emitter has successfully generated an entangled photon based on the detection time and the communication distance. Furthermore, when using a single emitter per node, the entanglement distribution rate over a distance $L$ is limited by the classical communication time $L/c$. In contrast, employing a large register linearly increases this rate by a factor of $n$. Fig. 5A displays $\eta_{overall}$ and thus demonstrates the expected increase in efficiency as we add atoms to the register. We measure a maximum detection efficiency of $\qty{88.6(1)}{\percent}$ for $n=6$ atoms, and extract the in-fiber photon generation efficiency $\eta_{fiber}$ by correcting $\eta_{overall}$ for the finite propagation and detection efficiencies (\qty{70(7)}{\percent}). As shown in Fig. 5A, we find $\eta_{fiber}=\qty{48(5)}{\percent}$ for a single atom, increasing up to \qty{97.4(6)}{\percent} for six atoms. The increase in efficiency translates directly into an almost linear increase in the average number of emitted photons $\overline{n}$ per attempt (Fig. 5B), demonstrating the increase in rate when using a register. We achieve $\overline{n}=1.88(1)$ for six atoms, more than a five-fold increase compared to the case of an individual atom.
\begin{figure}[htbp]
\centering
\includegraphics[width=\columnwidth]{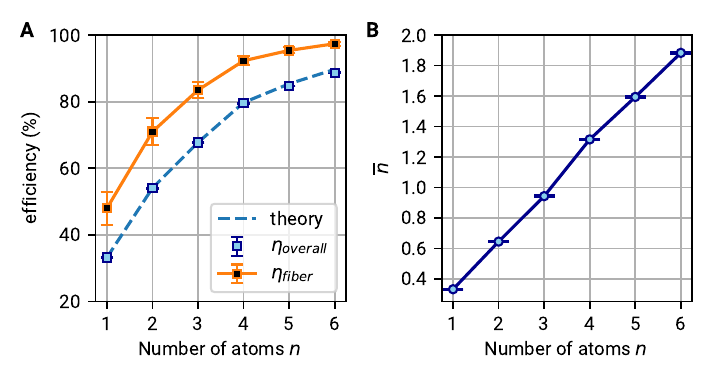}
\caption{\label{fig:efficiency} \textbf{Multiplexed efficiency and average photon number for an increasing number of atoms.} Efficiency to generate and fiber-couple ($\eta_{fiber}$), or to generate, fiber-couple and detect ($\eta_{overall}$), at least one entangled photon \textbf{(A)} and average photon number $\overline{n}$ \textbf{(B)} as a function of the number of atoms. The theory curve in part (A) is calculated using $\eta_{overall}= 1- \prod_{i=1}^n (1-\eta_i)$, with $\eta_i$ the probability to generate a photon from the $i$-th atom and the values $\eta_i$ as in Fig. 2. As a guide for the eye, the blue curve in (B) connects the data points. For six atoms, we measure an overall (in-fiber) photon detection efficiency of $\eta_{overall}=\qty{88.6(1)}{\percent}$ ($\eta_{fiber}=\qty{97.4(6)}{\percent}$) compared to \qty{33.2(2)}{\percent} (\qty{48(5)}{\percent}) for a single atom. We observe a nearly linear increase in $\overline{n}$ up to \qty{1.88(1)} photons per attempt.
}
\end{figure}

\quad \newline
\quad \newline
\textbf{Concluding remarks.} The atomic array assembling scheme reported here could boost the duty cycle of the recently demonstrated two-atom quantum-repeater node which had rather inefficient probabilistic loading \cite{langenfeld_quantum_2021}. Moreover, our system could potentially host many more atoms, as the two-dimensional array of tweezers could be extended to accommodate 40 atoms along the cavity axis, limited by the field of view of the addressing system (\qty{200}{\micro \metre}), and 7 atoms in the transversal direction, limited by the cavity waist. The latter number can be further increased by transporting atoms in and out of the cavity \cite{Nussmann2005,Bluvstein2022}, with atoms outside the cavity serving as a quantum reservoir, similar to what has been proposed for scalable quantum computation with ions \cite{Kielpinski2002}. With a spacing between atoms around or less than $\qty{10}{\micro \metre}$, smaller than a typical Rydberg blockade radius, intra-cavity quantum-logic operations between neighboring qubits seem possible \cite{Ramette2022}. Alternatively, the all-to-all connectivity of the atomic qubits in the cavity allows for quantum information processing between large numbers of qubits in a single step \cite{Duan_2005,Lin_2006,welte_photon-mediated_2018}, in comparison to all hitherto implementations of quantum circuits where these gates are typically decomposed into two-qubit gates and single-qubit rotations \cite{Shende2008}.

\section*{Acknowledgments}
\textbf{Funding:}
This work was supported by the Bundesministerium f\"{u}r Bildung und Forschung (QR.X, Förderkennzeichen 16KISQ019), by the Deutsche 
Forschungsgemeinschaft under Germany’s Excellence Strategy (MCQST, 
Project No. 390814868), and by the European Union’s Horizon Europe research and innovation program via the project Quantum Internet Alliance (QIA, Grant Agreement No. 101102140). E.D. acknowledges support from the Max Planck-Harvard Research Center for Quantum Optics postdoctoral scholarship. S.W. acknowledges financial support from the SNSF Swiss Postdoctoral Fellowship (Project no. TMPFP2$\_$210584). \textbf{Authors contributions:} All authors contributed to the experiment, the analysis of the results, and the writing of the manuscript. \textbf{Competing interests:} The authors declare that there are no competing financial interests. \textbf{Data availability:} The data underlying the figures are deposited at Zenodo \cite{Repository}. All other data needed to evaluate the conclusions in the paper are present in the main text or the Supplementary Material.

\newpage
\quad \newpage
\section*{Supplementary material}
\setcounter{equation}{0}
\subsection{Preparation of tweezer arrays and addressing}
At the beginning of the experimental sequence, we trap \isotope[87]{Rb}{} atoms in a magneto-optical trap (MOT) outside the resonator. After the MOT phase, we load the atoms into a red detuned dipole trap ($\lambda = \qty{1064}{\nano \metre}$) to transport the atoms over a distance of \qty{14}{\milli \metre} from the position of the MOT to the center of the cavity. For this purpose, the dipole trap is focused halfway between the resonator and the position of the MOT, creating an optical potential in which the atoms can freely oscillate between the MOT region and the resonator region. After traveling through the potential exactly once, the atoms reach the center of the resonator. This half-period oscillation takes \qty{70}{\milli \second}. Meanwhile, before the atoms reach the resonator, we turn on a two-dimensional array of optical tweezers midway between the two resonator mirrors. Each individual tweezer has a waist of \qty{1.40(5)}{\micro \meter} and a trap depth of about \qty{1}{\milli \kelvin} which corresponds to a power of \qtyrange{1.5}{2}{\milli \watt} per individual tweezer, resulting in an atomic lifetime of about \qty{20}{\second} in the tweezers. Since we are only using the tweezers to prepare the atomic arrays, we use a rather near-resonant wavelength of \qty{797}{\nano \metre}, as here little power is needed for the desired trap depth. Once the atoms reach the center of the cavity in the transport beam, they move across and are stochastically captured in the tweezers array in the presence of appropriate laser cooling. We use the scattered cooling light to image the atoms and assess the state of the preparation. Based on the image evaluation, another set of tweezers, the moving tweezers, controlled by the second AOD, arranges the atoms into the desired shape of the array. The arrangement of the atoms takes another \qtyrange{1}{2}{\second}, as we arrange the atoms sequentially with a moving speed of \qty{5}{\micro \metre / \milli \second}, because using a faster moving speed leads to atom loss. The arrangement protocol is implemented to minimize the distance and number of moves, and in principle the arrangement procedure in a cavity environment is the same as in a free space experiment once the atoms are trapped in the initial static array. We track the progress of the atom rearrangement and when it is complete, we load the atoms into two standing wave traps at \qty{1064}{\nano \metre} and \qty{770} {\nano \metre}, respectively. Here, the atomic storage time is $\sim \qty{1}{\minute}$. Once this step is completed, we start the experimental science sequence. The overall probability of success for this preparation protocol depends on the number of atoms. It is as high as \qty{90}{\percent} for two atoms and \qty{20}{\percent} for an array of six atoms. In our previous stochastic loading scheme without optical tweezers, we immediately transferred the atoms to the optical lattice after transporting them in the transport trap. In this way of loading atoms, the number of atoms in the resonator (the average number of atoms loaded could only be influenced by changing the size of the MOT) and the distance between the atoms are stochastic. Therefore, the probability of loading three atoms at the correct distance from each other is already less than \qty{0.5}{\percent} and working with larger arrays was not possible due to the low probability of success. Extrapolating from the measured values for the old loading scheme, we estimate that for six atoms the probability of success is about four orders of magnitude higher when the atoms are loaded with the tweezers. The individual addressing of the atoms is controlled by the same AOD as the moving tweezers. We observe atom-hopping between different lattice sites along the cavity direction at a rate of $<\qty{1}{\Hz}$. We attribute this to heating caused by the fluctuations of the intra-cavity $\qty{770}{\nano \meter}$ standing-wave intensity, resulting from the fluctuations of the cavity length. We overcome this by continuously monitoring the position of the atoms at a rate of $\qty{3}{\Hz}$ and adjusting the AODs frequency accordingly. As the generation-to-detection efficiency is at the theoretical maximum, this shows that hopping is not a limitation at present. In the future, we plan to further reduce the hopping rate by ground-state cooling the atom \cite{ReisererGroundState2013}. We also plan to implement a faster atom-position detection by increasing our current imaging setup signal-to-noise ratio.
\subsection{Detailed level scheme}
A detailed level scheme of the employed \isotope[87]{Rb} atoms is shown in Fig. S1. The two hyperfine ground states $\ket{F{=}1}$ and $\ket{F{=}2}$ are split by a microwave transition at \qty{6.8}{\giga \Hz}. The cavity is detuned from the $\ket{F{=}1}\rightarrow \ket{F^\prime{=}1,m_F{=}-1}$ transition by $\Delta_{ac}=2\pi{\times} \qty{200}{\mega \Hz}$.
\renewcommand{\thefigure}{S1}
\begin{figure}[]
\centering
\includegraphics[width=\columnwidth]{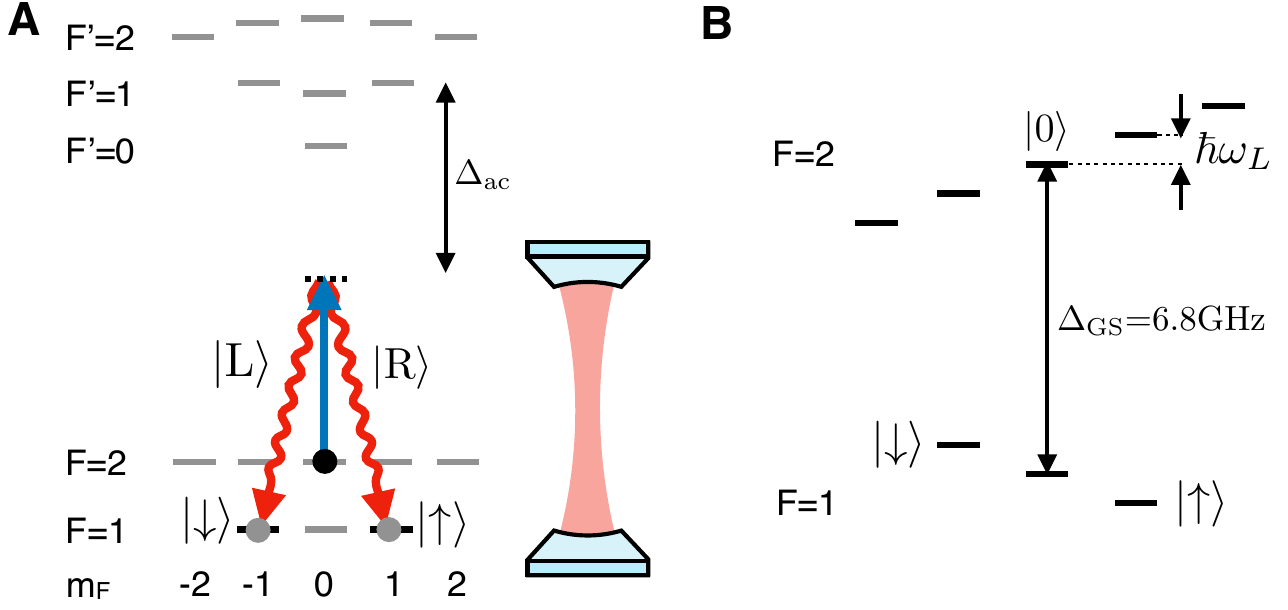}
\caption{\label{fig:levelScheme_supp} \textbf{Detailed level scheme of the employed \isotope[87]{Rb} atoms. (A)} At the start of the experiment, the atom is initialized in the state $\ket{F{=}2, m_F{=}0}$. A vSTIRAP pulse (blue arrow) via the addressing system triggers the release of a single photon (red wiggly arrows) into the cavity mode. The cavity, shown schematically on the right, is tuned, such that it has a detuning of $\Delta_{ac}=2\pi\times \qty{200}{\mega \Hz}$ from the $\ket{F{=}1}\rightarrow \ket{F^\prime{=}1, m_F{=}-1}$ transition. \textbf{(B)} Level scheme of the hyperfine ground states $\ket{F{=}1}$ and $\ket{F{=}2}$. As we apply a guiding field during the experimental sequence, we induce a Zeeman splitting with a Larmor frequency of $\omega_L=2\pi \times \qty{100}{\kHz}$.}
\end{figure}

\subsection{Raman Transitions}
\label{sec:Raman}
 To drive transitions between the different Zeeman sublevels of the $\ket{F{=}1}$ and $\ket{F{=}2}$ manifold, we employ a stimulated Raman transition as illustrated in Fig. S2. The probe and coupling beam are from the same laser oscillator to ensure maximum phase stability. The frequency of the beams is set to be on two-photon resonance with the involved states, i.e. to $\omega_{\ket{\uparrow} \rightarrow \ket{F{=}2, m_F{=}2}}$ in Fig. S2, and the frequency splitting of \qty{6.8}{\giga \Hz} is generated with a combination of an EOM and AOMs. The wavelength of the Raman laser is $\lambda = \qty{788.24}{\nano \metre}$ to give minimum differential light shift on the individual atoms. The beams are impinging from the side along the $x$-axis, see Fig. 1A (main text), on the cavity with a waist of $w_0=\qty{82}{\micro \metre}$ and are polarized with $\sigma^+ + \sigma^-$ and $\pi$ polarization. This allows to drive transitions with $\Delta m_F{=}\pm1$. In Fig. S2, we illustrate the transition from $\ket{\uparrow} $ to $\ket{F{=}2,m_F{=}2}$, but since the probe beam is polarized with $\sigma^+ + \sigma^-$ polarization, we can also drive the transitions from $\ket{\downarrow} $ to $\ket{F{=}2,m_F{=}-2}$ or from $\ket{\uparrow}$ or $\ket{\downarrow}$ to $\ket{0}$ by adjusting the frequency of the control beam (and hence the two-photon frequency) as the individual $m_F$-states are separated in frequency, see Fig. S1.

\renewcommand{\thefigure}{S2}
\begin{figure}[]
\centering
\includegraphics[width = \columnwidth]{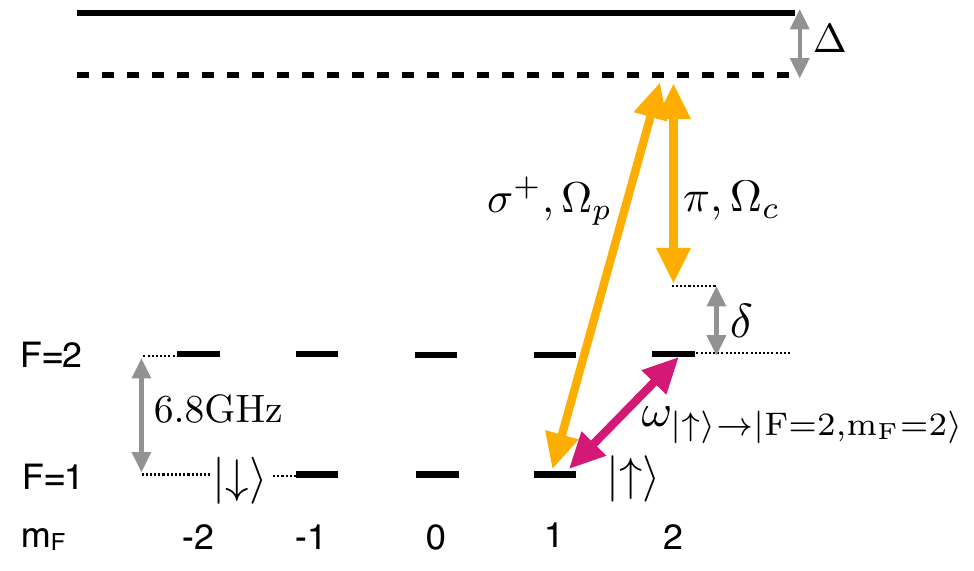}
\caption{\label{fig:supp_5}\textbf{Raman transition from $\ket{\uparrow}$ to $\ket{F{=}2, m_F{=}2}$.} 
The Raman transitions are driven with two Raman beams (yellow arrows) $\Omega_c$ and $\Omega_p$ which are $\pi$ and $\sigma^+ + \sigma^-$-polarized, respectively. These beams allow to bridge the hyperfine transition of $6.8$GHz. We show the transition from $\ket{\uparrow}$ to $\ket{F{=}2, m_F{=}2}$ with the single-photon detuning $\Delta$, the two-photon detuning $\delta$ and the two-photon frequency $\omega_{\ket{\uparrow}\rightarrow \ket{F{=}2, m_F{=}2}}$.} 
\end{figure}

\subsection{Experimental protocol}
\label{sec:CompProt}
The complete experimental sequence including the timing is shown in Fig. S3. It starts by preparing the atoms into the $\ket{5S_{1/2}, F{=}2,m_F{=}0}$ state. For the preparation, we use a combination of $\pi$-polarized light resonant with the $\ket{5S_{1/2}, F{=}2,m_F{=}0}\rightarrow\ket{5P_{1/2}, F^\prime{=}2,m_F{=}0}$ transition and a repumper resonant with the $\ket{5S_{1/2},F{=}1}\rightarrow \ket{5P_{3/2},F{=}2}$ transition. The pumping beams impinge on the atoms along the $x$-axis (see Fig. 1A (main text)) and initialize all atoms simultaneously with an efficiency of \qty{80(5)}{\percent} and a pumping fidelity of at least $\qty{98}{\percent}$ in $\ket{F{=}2,m_F{=}0}$. We estimate the lower limit of the fidelity by looking at the minimum of the correlations in the $ZZ$ basis, see section \ref{sec:FidelityMeasurement}. The bases are defined in table S1. 

\renewcommand{\thetable}{S1}
\begin{table}[htb] 
\begin{center}
\begin{tabular}{ | l| l|} 
\hline
Photonic bases & Atomic bases\\
 \hline & \\[-1.5ex]
 $\begin{array} {@{}c@{}c@{}c@{}c} X_p =\{ &\ket{H} &= \frac{1}{\sqrt{2}}\left( \ket{R}+ \ket{L}\right),\\ &\ket{V}& = \frac{1}{\sqrt{2}}\left( \ket{R}-\ket{L} \right)\}  \end{array}$& $\begin{array} {@{}c@{}c@{}c@{}c} X_a  =  \{ &\ket{\uparrow_x}& = \frac{1}{\sqrt{2}}\left( \ket{\uparrow}+\ket{\downarrow}\right), \\ &\ket{\downarrow_x}& = \frac{1}{\sqrt{2}}\left( \ket{\uparrow}-\ket{\downarrow} \right)\} \end{array}$\\ [+2.5ex]
 \hline& \\[-1.5ex]
$\begin{array} {@{}c@{}c@{}c@{}c} Y_p =\{ &\ket{A}& = \frac{1}{\sqrt{2}}\left( \ket{R}+i\ket{L} \right),\\ &\ket{D}& = \frac{1}{\sqrt{2}}\left( \ket{R}-i \ket{L}\right)\}\end{array}$ &$ \begin{array} {@{}c@{}c@{}c@{}c} Y_a =\{ &\ket{\uparrow_y}& = \frac{1}{\sqrt{2}}\left( \ket{\uparrow}+ i\ket{\downarrow}\right),\\ &\ket{\downarrow_y}& = \frac{1}{\sqrt{2}}\left( \ket{\uparrow}- i\ket{\downarrow} \right)\} \end{array} $\\ [+2.5ex]
 \hline & \\[-1.5ex]
$Z_p =\left\{ \ket{R} , \ket{L} \right\}$& $Z_a = \left\{ \ket{\uparrow}, \ket{\downarrow}\right\}$ \\[+0.5ex]
\hline
\end{tabular}
\end{center}
\caption{\label{tab:BasesDefinitions} \textbf{Definition of the atomic and photonic bases.} The table displays the definitions we use in equations (\ref{eq:PsiPlusDiffBases}) and (\ref{eq:PsiPlusDiffBases2}).}
\end{table}

Here and in the following, the terms $XX$, $YY$, and $ZZ$ basis stand for a combined atom-photon measurement along the basis $X_a \otimes X_p$, $Y_a 
\otimes Y_p$ and $Z_a \otimes Z_p$ respectively. \newline
\newline
\renewcommand{\thefigure}{S3}
\begin{figure*}[]
\centering
\includegraphics[scale=0.35]{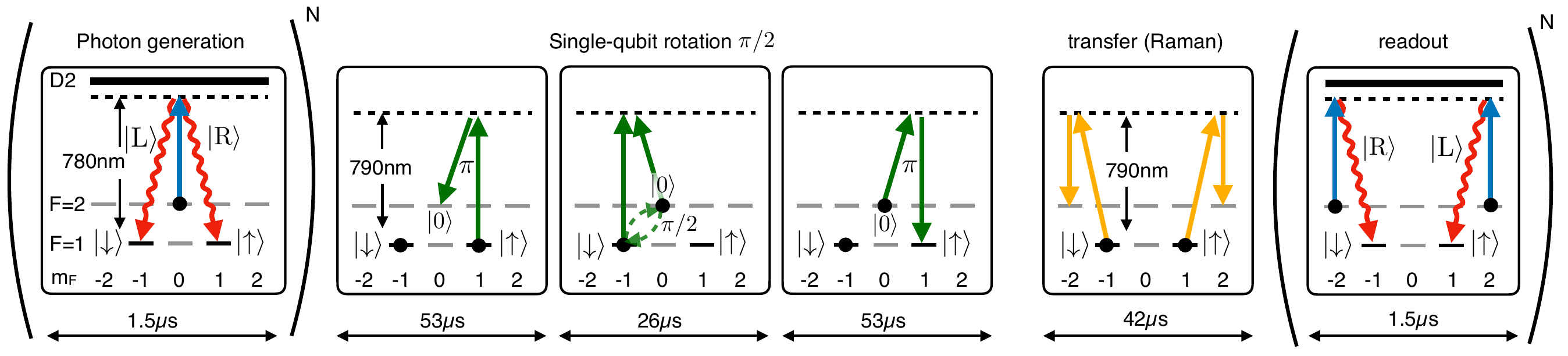}
\caption{\label{fig:ExpSequence_supp} \textbf{Complete experimental sequence.} We initially generate $N$ signal photons (red wiggly arrows) with vSTIRAP pulses (blue arrow) by individually addressing the $N$ intra-cavity atoms. Subsequently, we set the readout basis of the atomic spin using $\pi/2$ and $\pi$ pulses (green arrows). Both are addressing all atoms globally. Eventually, we map the information stored in the atomic spin to readout photons (red wiggly arrows) with two global Raman $\pi$ pulses (yellow arrows) and an additional set of N individual vSTIRAP pulses (blue arrows).}
\end{figure*}
Then, $N$ vSTIRAP pulses sequentially illuminate the atoms and ideally, we generate $N$ entangled atom-photon pairs. The vSTIRAP pulse and the resonance frequency of the cavity are set to have the same single-photon detuning with respect to the excited state $\ket{F^\prime=1, m_F{=}0}$ and the transitions from $\ket{F{=}2, m_F{=}0} \rightarrow \ket{F{=}1, m_F{=}1}$ and $\ket{F{=}2, m_F{=}0} \rightarrow \ket{F{=}1, m_F{=}-1}$ are on two-photon resonance (see Fig. S1). After the emission of the photon, the atoms are in the $\ket{F{=}1}$ ground state, and the photons emitted into the cavity mode can be scattered by other atoms in the same ground state. This scattering rate scales as $g^2/ \Delta_{ac}$ and therefore it is favorable to choose a large detuning $\Delta_{ac}$. However, choosing too large a detuning, i.e. $\left | \Delta_{ac}  \right | \gg H_{1^\prime, 3^\prime}$, with $H_{1^\prime, 3^\prime}$ being the hyperfine splitting between $\ket{F^\prime{=}1}$ and $\ket{F^\prime{=}3}$, increases the probability of exciting to $\ket{F^\prime{=}3}$, resulting in a reduction in fidelity \cite{langenfeld_quantum_2022}. It is therefore necessary to set the single photon detuning $\Delta_{ac}$ between the cavity and the atom carefully. We choose a detuning of $\Delta_{ac}=2\pi \times \qty{200}{\mega \Hz}$ with respect to the $\ket{F^\prime{=}1, m_F{=}-1}$ state. The rest of the sequence depends on the measurement basis. For the $Z_a$ basis state, we perform two $\pi$-pulses followed by another vSTIRAP pulse. The $\pi$ pulses map the state $\ket{\uparrow}$ to $\ket{F{=}2, m_F{=}2}$ and the state $\ket{\downarrow}$ to $\ket{F{=}2,m_F{=}-2}$. The consecutive vSTIRAP pulse generates a readout photon transferring the atoms back to the states $\ket{\uparrow}$ and $\ket{\downarrow}$. A polarization measurement of the readout photon along the $Z_p$ basis projects the atoms onto one specific state, explicitly a $\ket{L}$ detection projects to $\ket{\uparrow}$ while $\ket{R}$ to $ \ket{\downarrow}$. If we measure the atom along $X_a$ or $Y_a$, we introduce in the sequence additional Raman pulses to perform a rotation of an angle $\pi/2$ with a phase $\theta=0$ or $\theta=\pi/2$ in the atomic qubit space. This maps $\ket{\uparrow_x}\textbackslash \ket{\downarrow_x}$ and $\ket{\uparrow_y}\textbackslash \ket{\downarrow_y}$ onto $\ket{\uparrow}\textbackslash \ket{\downarrow}$, depending on the chosen phase. The $\pi/2$ rotation is composed of three Raman pulses:
\begin{enumerate}
\item A $\pi$ pulse to transfer the state $\ket{\uparrow}=\ket{F{=}1,m_F{=}+1}$ to $\ket{F{=}2,m_F{=}0}\equiv \ket{0}$. The qubit is thus stored in the $\{\ket{\downarrow}, \ket{0}\}$-manifold.
\item \label{Pulse2} A $\pi/2$ pulse between $\ket{\downarrow}$ and $\ket{0}$ with phase $\theta=0$ or $\theta=\pi/2$, depending on the measurement basis $X_a$ or $Y_a$ respectively, resulting in the map
\begin{eqnarray}
\frac{1}{\sqrt{2}}\left(\ket{\downarrow}+\ket{0}\right) &\rightarrow \ket{0} \label{eq:pi_2_1}\\
\frac{1}{\sqrt{2}}\left(\ket{\downarrow}-\ket{0}\right) &\rightarrow \ket{\downarrow} \label{eq:pi_2_2}\\
\frac{1}{\sqrt{2}}\left(\ket{\downarrow}+ i\ket{0}\right) &\rightarrow \ket{0}\label{eq:pi_2_3}\\
\frac{1}{\sqrt{2}}\left(\ket{\downarrow}- i\ket{0}\right) &\rightarrow \ket{\downarrow} \label{eq:pi_2_4}.
\end{eqnarray}
where Eqs. (\ref{eq:pi_2_1},\ref{eq:pi_2_2}) correspond to $\theta=0$ and Eqs. (\ref{eq:pi_2_3},\ref{eq:pi_2_4}) to $\theta=\pi/2$. 

\item An additional $\pi$ pulse to transfer the population from $\ket{0}$ back to $\ket{\uparrow}$. Afterwards, the qubit is again encoded in the $\{\ket{\downarrow}, \ket{\uparrow}\}$-manifold.
\end{enumerate}
\noindent These rotations, unlike the vSTIRAP pulses, are global pulses, i.e., they are executed on all atoms simultaneously. After the $\pi/2$ rotation, one proceeds again as in the $Z_a$-basis and reads out the atomic qubit by mapping the atomic state to the polarization with the scheme described above.

\noindent To set $\theta$ we note that the Raman transitions $\ket{\uparrow} \rightarrow \ket{0}$ and $\ket{0} \rightarrow \ket{\downarrow}$, have two different two-photon resonance frequencies $\omega_{\ket{\uparrow} \rightarrow \ket{0}} = \Delta + \omega_L$, and $\omega_{\ket{0} \rightarrow \ket{\downarrow}} = \Delta - \omega_L$ respectively, with $\Delta=2\pi\times\qty{6.8}{\giga \Hz}$ the splitting between the $\ket{F{=}1}$ and $\ket{F{=}2}$ hyperfine ground states (see Fig. S1B). This means that after the transfer of the population from $\ket{\uparrow}$ to $\ket{0}$ with pulse (1.), the qubit evolves at a frequency of $\omega_{\ket{\downarrow} \rightarrow \ket{0}}$ while the RF frequency driving the Raman laser is set to $\omega_{\ket{\uparrow} \rightarrow \ket{0}}$. To apply the pulse (2.), the RF frequency is switched to $\omega_{\ket{\downarrow} \rightarrow \ket{0}}$ after a time $T$ which results in an accumulated phase of $\theta=2\omega_L T$. A carefully chosen $T$ allows to set the phase $\theta$ at will. To determine $T$ corresponding to a measurement along the $X_a$ ($Y_a$) basis, we measure the signal photon along $X_p$ ($Y_p$), scan $T$, and measure the readout photon along $Z_p$. We then extract the probability $p(i|j)$ of finding the signal photon in the polarization $i=\{H,V\}$ ($i=\{D,A\}$) conditioned on the readout photon in $j=\{R,L\}$, see Fig. S4. $p(i|j)$ oscillates as a function of $\theta$, and a measurement along the $X_a$ ($Y_a$) basis corresponds to a $T$ where $p(H|L)$ ($p(D|L)$) is maximum.
\noindent For the polarization analysis of the signal photon in $XX$ and $YY$, we need to set the basis of the detection setup to a linear polarization. However, the readout photon is always detected in the $\{\ket{R},\ket{L}\}$ basis. Therefore, the polarization basis of the detection setup must be changed between the detection of the signal and the readout photons. This is achieved by using an EOM, which switches the basis between linear and circular polarization in less than $\qty{1}{\micro \second}$. To switch between the $XX$ and the $YY$ basis, a $\lambda/2$ waveplate (not shown in Fig. 1B) is inserted in front of the EOM to adjust the detection basis accordingly. \newline 

\renewcommand{\thefigure}{S4}
\begin{figure}[]
\centering
\includegraphics[scale = 1]{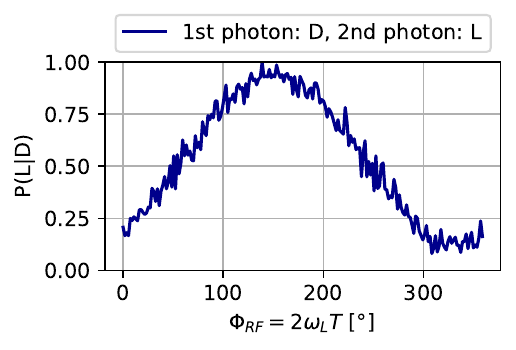}
\caption{\label{fig:PhaseScan_supp} \textbf{Phase scan of the $\pi/2$ pulse in the $YY$ basis.} To perform the $\pi/2$ rotation between $\ket{\uparrow}$ and $\ket{\downarrow}$, we first transfer the population from $\ket{\uparrow}$ to $\ket{0}$ (see S3). As the two photon detunings of the two Raman transitions $\omega_{\ket{\uparrow} \rightarrow \ket{0}}$ and $ \omega_{\ket{0} \rightarrow \ket{\downarrow}}$ differ by $2\times \omega_L$, the qubit in the $\left\{\ket{0}, \ket{\downarrow}\right\}$ manifold evolves at a frequency of $2 \times \omega_L$ with respect to the previously set frequency $\omega_{\ket{\uparrow} \rightarrow \ket{0}}$. Therefore, to set the phase for the $\pi/2$ pulse properly, we have to scan the time $T$ at which we switch the two-photon detuning from $\omega_{\ket{\uparrow} \rightarrow \ket{0}}$ to $ \omega_{\ket{0} \rightarrow \ket{\downarrow}}$. These scans are performed for both the $XX$ and $YY$ basis and we set the time $T$ to the maximal amplitude of the oscillatory signal.
}
\end{figure}
\subsection{Fidelity measurement}
\label{sec:FidelityMeasurement}
In this section, we describe how we determine the fidelity of the prepared atom-photon states with ideal Bell states. Ideally, after the photon emission we generate the state
\begin{equation}
\ket{\Psi^+} =\frac{1}{\sqrt{2}} \left( \ket{R, \uparrow}+\ket{L, \downarrow}\right).
\end{equation}
and similarly in the $XX$ and $YY$ basis: 
\begin{eqnarray}
\label{eq:PsiPlusDiffBases}
\ket{\Psi^+} &= \frac{1}{\sqrt{2}} \left( \ket{H, \uparrow_x}+ \ket{V, \downarrow_x}\right) \\
\label{eq:PsiPlusDiffBases2}
\ket{\Psi^+} &= \frac{1}{\sqrt{2}} \left( \ket{D, \uparrow_y}-\ket{A, \downarrow_y}\right).
\end{eqnarray}
We thus see that correlations between linear polarization and superposition states of $\ket{\uparrow}$ and $\ket{\downarrow}$ arise, e.g., between $\ket{H}$ and $\ket{\uparrow_x}$ or $\ket{D}$ and $\ket{ \uparrow_y}$.\newline
The fidelity is given by $\mathcal{F}=\mathrm{Tr}\left(\ket{\Psi^+}\bra{\Psi^+}\rho\right)$, which in turn can be simplified to $\mathcal{F}=\bra{\Psi^+}\rho\ket{\Psi^+}$. When writing the density operator in the basis of the Pauli matrices \linebreak $\rho = \sum_{k,j=0}^3 S_{k,j} \sigma_k \otimes \sigma_j$, we obtain the following expression for the fidelity 
\begin{equation}
\label{eq:FidDef_supp}
\mathcal{F} = \frac{1}{4}\left( 1+ S_{xx} + S_{yy} - S_{zz}\right).
\end{equation}
Here, $S_{xx}, S_{yy}$ and $S_{zz}$ are the two-qubit Stokes parameters. The individual two-qubit Stokes parameters are defined as \cite{Altepeter2004}
\begin{eqnarray}
S_{xx} &=& P_{\uparrow_x, \uparrow_x} - P_{\uparrow_x, \downarrow_x} - P_{\downarrow_x, \uparrow_x} + P_{\downarrow_x, \downarrow_x} \\
S_{yy} &=& P_{\uparrow_y, \uparrow_y} - P_{\uparrow_y, \downarrow_y} - P_{\downarrow_y, \uparrow_y} + P_{\downarrow_y, \downarrow_y}\\
\label{eq:StokesParams_supp}
S_{zz} &=& P_{\uparrow, \uparrow} - P_{\uparrow, \downarrow} - P_{\downarrow, \uparrow} + P_{\downarrow, \downarrow},
\end{eqnarray}
where $P_{jj}$ is the probability of measuring the state $\ket{j,j}$ and $P_{j,j} +P_{j,-j} + P_{-j,j}+ P_{-j,-j}=1$. Therefore, to determine the fidelity, we need to measure the correlations between the polarization of the signal photon and the internal atomic state in the three pairs of bases, as defined in table S1. From this, the fidelity can be calculated employing equations (\ref{eq:FidDef_supp}) to (\ref{eq:StokesParams_supp}). 

\subsection{Temporal shape of the photon}
\label{sec:PhotonShape}
The temporal shape of the photon is a source of infidelity in the superposition bases $XX$ and $YY$. After the detection of the photon, the atomic state is projected onto a superposition state of $\ket{\uparrow}$ and $\ket{\downarrow}$. As these states are energetically separated by $2 \hbar \omega_L $, the phase of the superposition states evolves in time. For example, in the $XX$ basis after the detection of the signal photon in $\ket{H}$, the atom is projected onto $\ket{\uparrow_x}$ which evolves in time as $\ket{\uparrow_x}(t)=\ket{\uparrow} + \exp\left(-2i \omega_L t) \ket{\downarrow}\right)$. Here, $t$ is the time that passed since the detection of the signal photon. As the temporal shape of the signal photon (see Fig. S5) has a non-vanishing width, the phase evolution of the atomic qubit after detection of the signal photon differs from shot to shot. Therefore, to reduce the influence of the photon shape on the fidelity, we define a time window $\tau$ and only consider events for which we have detected the photon within the specified time window. In the presented data we set the time window to $\tau=\qty{1.25}{\micro \second}$ which reduces the detection efficiency.

\renewcommand{\thefigure}{S5}
\begin{figure}[]
\centering
\includegraphics[scale = 1]{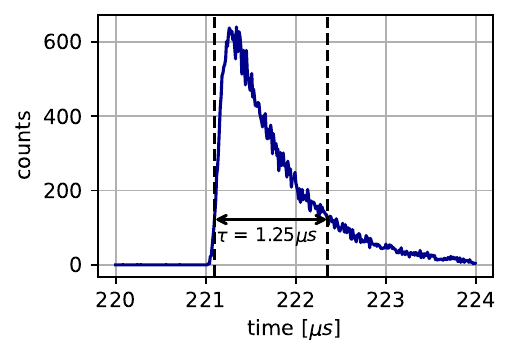}
\caption{\label{fig:PhotonShape_supp}\textbf{Typical temporal shape of a generated photon.} The finite temporal width of the photon reduces the fidelity of the produced entangled atom-photon states with ideal Bell states. To minimize this effect, we introduce a photon acceptance window of $\tau= \qty{1.25}{\micro \second}$. Photons outside of this interval are not used. This reduces the photon generation efficiency, i.e. in the illustrated case by almost $\qty{15}{\percent}$.
} 
\end{figure}

\subsection{Analysis of experimental imperfections}
In this section, we elucidate the various sources of errors in our experiment that contribute to inaccuracies in the entanglement generation. One major source of error stems from the temporal mode of the photon, see section \ref{sec:PhotonShape}. We calculate the influence of the temporal shape on the fidelity for a photon acceptance window of $\tau = \qty{1.25}{\micro \second}$ and a Larmor frequency of $\omega_L= 2\pi \times \qty{100}{\kilo \Hz}$ and obtain a maximum fidelity of $\mathcal{F}=\qty{96.2}{\percent}$. This contribution could be reduced by either a smaller photon acceptance window $\tau$ (resulting in a reduction of the efficiency), a smaller Larmor frequency (resulting in a smaller Rabi-frequency), or by increasing the power in the vSTIRAP. The latter, however, leads to atomic heating, reducing the storage time and the increasing tunneling to different lattice sites, which results in a lower overall efficiency. Another contribution comes from the atom preparation and state readout for which we estimate an infidelity of \qty{2.7(2)}{\percent}. The coherence time is \qty{1.1}{\milli \second} in the $\left\{\ket{\uparrow}, \ket{\downarrow}\right\}$-manifold. As the duration of the sequence depends on the number of atoms that we use, the influence of the coherence time is the largest for the six-atom case, where we attribute approximately \qty{4.5}{\percent} (compared to \qty{3}{\percent} for the two-atom case) of infidelity to the limited coherence. We attribute the remaining \qty{2.5}{\percent} of infidelity to the $\pi/2$ rotation between $\ket{\uparrow}$ and $\ket{\downarrow}$. We estimate the errors given by the polarization settings and the polarization of the vSTIRAP pulse to be below \qty{1}{\percent}.

\subsection{Coherence time scaling analysis}
The measured coherence time for the qubit states $\ket{\uparrow}$ and $\ket{\downarrow}$ is $\sim \qty{1.1}{\milli \second}$. With this coherence time, we see a decrease in the fidelity of the first (last) addressed atom of \qty{4.5}{\percent} (\qty{1.5}{\percent}) for six atoms. However, using dynamical decoupling as in \cite{langenfeld_quantum_2021_supplement}, the coherence time can be extended to $\sim \qty{20}{\milli \second}$, which allows us to increase the number of atoms in the register without appreciable decoherence. For example, for 100 atoms, a maximum fidelity of $\sim \qty{96}{\percent}$ for the atom that was addressed first, could be expected. This assumes a worst-case scenario of exponential decay with a coherence time of \qty{20}{\milli \second} and considers that \qty{15}{\micro \second} are needed to generate a photon and adjust the input RF frequency of the AODs accordingly to address the next atom. By transferring the atomic qubit into decoherence-protected superpositions, the coherence time can be further extended, resulting in a coherence time of $\qty{100}{ms}$ \cite{Koerber2013}. This would further increase the maximum fidelity to $\sim \qty{99.2}{\percent}$ for the atom that was addressed first.

\end{document}